\documentclass[useAMS,usenatbib]{mn2e}
\usepackage{epsfig}
\usepackage{times}
\usepackage{subeqn}
\usepackage{amssymb}

\def \implies {\Rightarrow}

\def \grad {\nabla} \def \del {\nabla}
\def \centreline {\centerline}
    
\def \beqn {\begin{equation}}
\def \eeqn {\end{equation}}
\def \bdm {\begin{displaymath}}
\def \edm {\end{displaymath}}

\newcommand{\vect}[1]{\mathbf{#1}}

\newcommand{\E}[1]{\times 10^{#1}}
\def \pc {\,\mathrm{pc}} \def \kpc {\,\mathrm{kpc}} \def \Mpc {\,\mathrm{Mpc}} 
\def \kmpersec {\,\mathrm{km}\,\mathrm{s}^{-1}}
\def \Msun {M_{\odot}}

\def \d {\:\mathrm{d}}

\def \half {\frac{1}{2}}

\newcommand{\text}[1] {\; \textrm{#1} \;}

\newcommand{\eqref}[1]{(\ref{#1})}

\newenvironment{remark*}{\textit{Remark.}}{}

\def\spose#1{\hbox to 0pt{#1\hss}}
\def\simlt{\mathrel{\spose{\lower 3pt\hbox{$\mathchar"218$}}
     \raise 2.0pt\hbox{$\mathchar"13C$}}}
\def\simgt{\mathrel{\spose{\lower 3pt\hbox{$\mathchar"218$}}
     \raise 2.0pt\hbox{$\mathchar"13E$}}}

\def \Sigmacrit {\Sigma_{\mathrm{cr}}}

\def \zl {z_{\mathrm{l}}}  \def \zs {z_{\mathrm{s}}}

\def \rhat {\hat{r}} \def \Rhat {\hat{R}}

\def \Ds {D_{\mathrm{s}}} \def \Dl {D_{\mathrm{l}}} \def \Dls {D_{\mathrm{ls}}}

\def \Ds {D_{\mathrm{s}}} \def \Dl {D_{\mathrm{l}}} \def \Dls {D_{\mathrm{ls}}}

\def \md {m_{\mathrm{d}}}  
\def \rhoc {\rho_{\mathrm{c}}} 
\def \Md {M_{\mathrm{d}}}
\def \Mb {M_{\mathrm{b}}} \def \rb {r_{\mathrm{b}}}
 \def \rh {r_{\mathrm{h}}}
\def \Mt {M_{\mathrm{t}}} \def \rs {r_{\mathrm{s}}} \def \rt {r_{\mathrm{t}}} \def \rthat {\hat{r_{\mathrm{t}}}}
\def \rhobar {\bar{\rho}}   

 \def \psid {\psi_{\mathrm{d}}} 

\def \Er {E_{\mathrm{r}}}  \def \rc {r_{\mathrm{c}}}   \def \rp {r_{\mathrm{p}}}  
\def \rd {r_{\mathrm{d}}}  \def \Mp {M_{\mathrm{p}}}  \def \Mdw {M_{\mathrm{sat}}}  \def \reff {r_{\mathrm{eff}}}
\def \Rfifty {d_{1/2}}   \def \Msat {M_{\mathrm{sat}}}

\title[PIEP_WithSatellites]
{The Effect of Satellite Galaxies on Gravitational Lensing Flux Ratios}

\author[E.M. Shin and N.W. Evans] {E.M. Shin\thanks{E-mail:
ems@ast.cam.ac.uk; nwe@ast.cam.ac.uk} and
N.W. Evans\footnotemark[1]\\ Institute of Astronomy, University of
Cambridge, Madingley Road, Cambridge, CB3 0HA, United Kingdom}

\begin{document}

\pagerange{\pageref{firstpage}--\pageref{lastpage}} \pubyear{2007}

\maketitle
\label{firstpage}

\begin{abstract}
Gravitational lenses with anomalous flux ratios are often cited as
possible evidence for dark matter satellites predicted by simulations
of hierarchical merging in cold dark matter cosmogonies. We show that
the fraction of quads with anomalous flux ratios depends primarily on
the total mass and spatial extent of the satellites, and the
characteristic lengthscale $\Rfifty$ of their distribution.  If
$\Rfifty \sim100\kpc $, then for a moderately elliptical galaxy with
a line-of-sight velocity dispersion of $\sim250 \kmpersec$, a mass of
$\sim3\E{9} \Msun$ in highly-concentrated (Plummer model) satellites
is needed for 20\% of quadruplets to show anomalous flux ratios,
rising to $\sim1.25 \E{10} \Msun$ for 50\%. Several times these
masses are required if the satellites have more extended Hernquist
profiles. Compared to a typical elliptical, the flux ratios of quads
formed by typical edge-on disc galaxies with maximum discs are
significantly less susceptible to changes through substructure --
three times the mass in satellite galaxies is needed to affect 50\% of
the systems.

In many of the lens systems with anomalous flux ratios, there is
evidence for visible satellites (e.g., B2045+265 or MG0414+0534). We
show that if the anomaly is produced by substructure with properties
similar to the simulations, then optically identified substructure
should not be preponderant among lens systems with anomalies. There
seem to be two possible resolutions of this difficulty. First, in some
cases, visible substructure may be projected within or close to the
Einstein radius and wrongly ascribed as the culprit, whereas dark
matter substructure is causing the flux anomaly. Second, bright
satellites, in which baryon cooling and condensation has taken place,
may have higher central densities than dark satellites, rendering them
more efficient at causing flux anomalies.
\end{abstract}


\section{Introduction}

The abundance of substructure in galaxy halos is emerging as a key
test in theories of galaxy assembly.  In Cold Dark Matter cosmogonies,
dark matter overdensities collapse to form cusped halos, with the
smallest and least massive halos being the densest. The simulations of
\citet{Kl99} and \citet{Mo99} predicted hundreds of small Galactic
satellite halos, in contrast to the nine then known satellites around
the Milky Way. \citet{Ef92} had already suggested that photoionisation
may lengthen the cooling times of gas in haloes with low circular
speeds. This effect suppresses the formation of satellite galaxies, but
produces a large population of entirely dark satellites~\citep[see
e.g.][]{Kr04,Mo06}.

In strong lensing, it has been known for some years that simple,
smooth models of galaxy lenses usually fitted the image positions
well, but the flux ratios of the images poorly. In a bold paper,
\citet{Da02} argued that the flux anomalies in gravitational lens
systems could be interpreted as evidence for entirely dark
substructures. They used 7 of the then available four-image lens
systems to claim detection of substructure amounting to a total mass
of $0.6 \% - 6 \%$ of the lens galaxy mass. However, this
interpretation was challenged as flux anomalies could arise from
alternative sources, such as absorption, scattering, or scintillation
by the interstellar medium of the lens galaxy, or by higher order
harmonics in the ellipsoidal models used to fit lens systems, or
stellar microlensing~\citep[see e.g.][]{Ev03,Ko04,Ma04}. In
particular, the surface mass density in substructure as judged from
simulations seems to be lower than that required by gravitational
lensing, at least within the Einstein radius which probes primarily
the inner parts of halos~\citep[e.g.,][]{Ma04}. It is also hard to
reproduce the observed statistics on cusp violations with substructure
~\citep{Ma06}. This argues against substructure as a primary cause of
anomalous flux ratios.

Nonetheless, there is good evidence in favour of substructure, and, in
some cases, visible substructure can be identified.  One of the
gravitational lens systems with a flux ratio anomaly is the radio-loud
quadruple CLASS B2045+265 discovered by \citet{Fa99}. Recent deep {\it
  Hubble Space Telescope} and {\it Keck} imaging of this system by
\citet{Mc07} have revealed the presence of a tidally disrupted dwarf
galaxy G2.  This may be the cause of the flux ratio anomaly, although
caution is needed as modelling suggests that G2 must be very highly
flattened ($q = 0.13$).  There is also evidence for visible structure
-- possibly a small galaxy -- in the radio-loud quadruple
MG0414+0534. \citet{Sc93} already argued that the perturbation caused
by this object may account for the relatively poor agreement between
the observed data on this lens and the theoretical models. The
quadruple lens systems CLASS B1608+656 \citep[see e.g.,][]{Fa96} has a
loose group of galaxies at the same redshift as the main lensing
galaxy, a phenomenon that also occurs in the six-image radio-loud
system B1359+154~\citep{Ru01}. It seems that visible substructure may
well be responsible for some of the flux ratio anomalies.

In addition, there have been striking observational developments
nearer to home. The last two years have seen the discovery of 10
faint, new Milky Way satellites in data from the Sloan Digital Sky
Survey (SDSS, see Willman et al. 2005, Zucker et al. 2006a,b;
Belokurov et al. 2006, 2007).  It seems likely that a population of
ultra-faint, dwarf galaxies does surround the Milky Way.  These may be
representatives of the ``missing satellites'', as predicted by the
simulations of \citet{Kl99} and \citet{Mo99}, or they may be a
population of tidal dwarf galaxies or even star clusters~\citep[see
e.g.,][]{Be07}. Such ultra-faint objects are only detectable nearby,
and so would be -- to all intents and purposes -- dark at the typical
redshifts of strong lenses.

All this suggests that it is worth re-examining the effects of
substructure on strong lenses.  In this paper, we answer the
following questions.  Given what we know about the satellite
populations, how frequently might we expect anomalous flux ratios for
elliptical and spiral galaxies?  If luminous satellite galaxies
represent the bright and massive end of a predominantly faint
population of objects, how frequently might we expect to attribute
flux ratio anomalies to visible objects? Three (B2045+265,
MG0414+0534, B1608+656) out of the sample of six quadruplets
originally identified by \citet{Da02} as anomalous have optically
identified companions that are possible causes. At the outset, this seems
surprisingly high, if the substructure is predominantly dark.

The paper is arranged as follows. In \S 2, models of elliptical and
spiral galaxies are briefly introduced, together with their satellite
galaxy populations. In \S 3, the frequency with which anomalous flux
ratios occur is shown to depend primarily on the scalelength of the
satellite distribution, the mass model used for the satellites and the
total mass in substructure, while depending only weakly on how the
mass is distributed between satellites. The simulations reported in \S
4 give the expected fraction of anomalous flux ratios for both
ellipticals and spiral lenses, together with the typical numbers
caused by high mass and luminous satellite galaxies.

\section{Methodology}

\subsection{Mass Models}

For convenience, we follow \citet{Sc92} by defining in the lens plane
the dimensionless distance, dimensionless surface mass density and
critical surface mass density
\begin{equation}\label{eq:2-DefDimLess-rAndkappaAndsigmacrit}
\vect{R} = \frac{\vect{\Rhat}}{\xi_{0}}, \qquad \kappa(\vect{R}) = \frac{\Sigma(\xi_{0}
  \vect{R})}{\Sigmacrit}, \qquad \Sigmacrit = \frac{c^{2}\Ds}{4 \pi G \Dl \Dls} \;,
\end{equation}
with $\Ds,\; \Dl,\; \Dls$ being the distances to the source, lens, and
between lens and source, and $\xi_{0}$ an arbitrary scale length which
relates the scaled (uncapped) coordinates to the unscaled
(capped). The corresponding dimensionless deflection potential is
the solution of the Poisson Equation
\begin{equation}
\kappa(x,y) = \frac{1}{2}\del^{2}\psi
\end{equation}
and the dimensionless bending angle is $ \mathbf{\alpha} = \grad\psi.$

\subsubsection{Primary Lens}

We examine two models for the main lens galaxy. The first is
appropriate for an elliptical galaxy lens. It is a pseudo-isothermal
elliptic deflection potential \citep[see e.g.][]{KK93,Hu01,Ev02}
\begin{equation}\label{eq:2-DefPIEP}
\psi(x,y) = \Er \left( \rc^{2} + (1-\epsilon)x^{2} + (1+\epsilon)y^{2}
\right)^{1/2}
\end{equation}
where $\rc$ is a dimensionless core radius (the length scale being the
arbitrarily chosen $\xi_{0}$), and
\begin{equation}\label{eq:wyn}
\Er = \sigma^{2}(\xi_{0} G \Sigmacrit)^{-1}
\end{equation}
is the dimensionless Einstein-ring radius of the singular isothermal
sphere with line-of-sight velocity dispersion $\sigma$ corresponding
to the $\rc = 0, \; \epsilon = 0$ case. There are two critical curves:
a small inner one which maps to a `radial' caustic and an outer
`tangential' one which maps to an astroid caustic. At the ranges of
$\epsilon$ we consider, the astroid caustic is wholly within the outer
caustic. A point source has one image if it is outside both caustics,
three if it is inside the outer caustic, and five if it is inside the
astroid caustic. Triplets and quintuplets, however, are effectively
doublets and quadruplets, because one of the multiple images is a
highly demagnified central image within the small inner critical
curve.  For concreteness, we consider the elliptical potential to have
(unless otherwise specified) a velocity dispersion $\sigma = 250
\kmpersec$ and a core radius of $100\pc$, as suggested by \cite{KK93}
and \cite{Ev02}. We restrict the ellipticity parameter to be smaller
than $\epsilon \approx 0.2$, otherwise the corresponding surface mass
density becomes dumbbell-shaped, which is inappropriate for elliptical
galaxies \citep[see][]{KK93}.

The second model is appropriate for a spiral galaxy lens. It is a
three-component model widely used in galactic astronomy (see
e.g. \citealt{Di99}, \citealt{Fe06}) as a model for the Milky Way. It
has a \citet{He90} bulge, a Miyamoto-Nagai (1975) disc and a cored
isothermal halo. The Hernquist bulge has 3D mass distribution
\begin{equation}\label{eq:MW-HernquistDensity}
\rho(\rhat) = \frac{\Mb}{2\pi}\:\frac{\rb}{\rhat (\rhat + \rb)^{3}} \:,
\end{equation}
where $\rhat$ is the spherical polar radius and $\rb$ a core
radius. That of the halo is
\begin{equation}\label{eq:MW-CISDensity}
\rho(\rhat) = \frac{\rhoc}{1 + \rhat^{2} / \rh^{2}} \:.
\end{equation}
where $\rh$ is the core radius and $\rhoc$ the central density. The
mass distribution of the Miyamoto-Nagai disc is complicated, but the
deflection potential in the edge-on case is simple:
\begin{equation}\label{eq:MW-MN-DefPot}
\psid = \half\md \,\log\left[ x^{2} + (a + \sqrt{b^{2} + y^{2}})^{2} \right]\:,
\end{equation}
where $\md = \Md / (\pi\Sigmacrit\xi_{0}^{2})$ is the dimensionless
mass and $a$ and $b$ control the shape of the distribution.  We
normalize our disc galaxy lens to the Milky Way, according to the
parameters given in \citet{Sh07}.
%
%
Disc galaxies give rise to three main different classes of
multiple-image configurations (see e.g. \citealt{Mo98}): `core
triplets' (in effect, doublets), `disc triplets' (where images straddle
the plane of the disc) and quintuplets (in effect, quadruplets). The
small 7-imaging butterfly cusp in the caustic of this edge-on Milky
Way is ignored here (see e.g. \citealt{Sh07}).

We choose the redshift of the lens to be $0.46$ and that of the source to
be $2.15$. These are the median redshifts of known 4-image lens
systems (see the CASTLES website), omitting those known to have more
than one main lens and those without known lens and source redshifts.
We use a flat $\Lambda$CDM cosmology with $\Omega_m = 0.27$,
$\Omega_\Lambda = 0.73$, $H_{0} = 71 \kmpersec\Mpc^{-1}$.

\subsubsection{Plummer Model Satellites}

The Plummer model is often fitted to observed dwarf spheroidal
galaxies (see e.g. \citealt{McC06}, \citealt{Wi02}). We give our
satellites densities
\begin{equation}\label{eq:2-DefPlummer-kappa}
\kappa(r_{k}) = \kappa_{0}^{(k)} \left( 1 + \lambda_{k}^{2} r_{k}^{2} \right)^{-2}
\end{equation}
where $r_{k} = \sqrt{ x_{k}^{2} + y_{k}^{2} }\:,$ $x_{k}$ and $y_{k}$
being Cartesian coordinates with their origin at the centre of the
$k$th satellite galaxy, $\kappa_{0}^{(k)}$ is the central density of
that galaxy, and $\lambda_{k} = \xi_{0} / \rp^{(k)}$ where $\rp^{(k)}$
is the Plummer model scale radius. The mass of any satellite is
\begin{equation}
\Mp^{(k)} = \Sigmacrit \kappa_{0}^{(k)} \pi \rp^{(k){2}} \:.
\end{equation}
We give a `typical' $10^{7}\Msun$ Plummer satellite a scale radius
(equal to its half-light radius) of $140\pc,$ which is the median
half-light radius of known Milky Way satellite spheroidal galaxies
(see e.g. \citealt{Be07}), and vary $\rp$ as $\sqrt{\Mp}$. This
mass-radius scaling is consistent with those found in $N$-body
simulations, where the size of satellites scales as a power of the
mass. We also put a lower limit on $\rp$ of $70\pc$, the smallest
known half-light radius of a Milky-Way dwarf spheroidal~\citep{Be07}.

\subsubsection{Hernquist Model Satellites}

We also examine a second density profile for the satellites, motivated
by numerical simulations. The NFW density profile~\citep{Na96}
\begin{equation}\label{eq:2-NFW-rho}
\rho_{\mathrm{NFW}} = \rho_{s} \left(\frac{\rhat}{\rs}\right)^{-1}\left(1 + \frac{\rhat}{\rs}\right)^{-2},
\end{equation}
is found to be a good fit to cold dark matter subhalos.  Rather than
using the NFW, which falls off as $\rhat^{-3}$ at large radii and
therefore has formally infinite total mass, we use the Hernquist
density profile, which is similar to the NFW profile, but has an
$\rhat^{-4}$ asymptotic density decay. Our Hernquist satellites are
less concentrated than the Plummers, and we make a lowest order
estimate of the effects of tidal stripping by truncating them at tidal
radii $\rthat^{(k)}$ defined by the condition
\begin{equation}\label{eq:2-SHQ-Defn-of-rt}
\rhobar^{(k)}(\rthat^{(k)}) = \rhobar(\rhat)
\end{equation}
where $\rhat$ is the radial distance of the $k$th satellite from the
centre of the main galaxy, $\rhobar^{(k)}(\rthat^{(k)})$ its mean mass
density, and $\rhobar(\rhat)$ the mean mass density due to the main
galaxy enclosed within a sphere of radius $\rhat$ when the
pseudo-isothermal elliptical potential is approximated by a singular
isothermal sphere, that is,
\begin{equation}\label{eq:2-SHQ-rhobar}
\rhobar(\rhat) = \left(\frac{4}{3} \pi \rhat^{3} \right)^{-1} \int_{0}^{\rhat} 4\pi\rhat^{\prime^{2}} \frac{\sigma^{2}}{2\pi G \rhat^{\prime^{2}}} \d \rhat^{\prime} \; = \; \frac{3 \sigma^{2}}{2\pi G} \: \rhat^{-2} \:.
\end{equation}
The truncated Hernquist profile is then
\begin{equation}\label{eq:2-SHQ-rho}
\rho(\rhat_k) = \left\{ \begin{array}{ll}
                      \frac{M^{(k)}}{2\pi\rs^{(k)^{2}}\rhat_{k}}(1 + \rhat_{k}/\rs^{(k)})^{-3}, & \rhat < \rthat^{(k)}  \\
                      0 \;, & \rhat > \rthat^{(k)} \:,
                    \end{array}
 \right.
\end{equation}
where $\rhat_{k}$ is the radial distance from the centre of the $k$th
satellite, $M^{(k)}$ are the untruncated Hernquist masses and
$\rs^{(k)}$ are Hernquist scale lengths.

The scale length of each satellite is chosen so that the location of
the peak circular velocity of the model (which is $\rs$) varies with
the mass bound within the tidal radius $\Mt \equiv M \rt^{2}/(\rt^{2}
+ \rs^{2})$ as
\begin{equation}\label{eq:2-SHQ-rs-Mt-scaling}
\rs = 10^{-4} \: \sqrt{\Mt / \Msun} \: \kpc
\end{equation}
in agreement with the numerical simulations of, for example, Diemand,
Kuhlen \& Madau (2007).

If the tidal radius of any Hernquist satellite is less than its scale
radius, we reject it as being too strongly tidally disrupted to be
approximated by this mass profile. For example, \citet{Me01} find that
NFW clumps lose their mass rapidly for $\rt < \rs$. It is beyond the
scope of this paper to consider lensing effects due to the
tidally-stripped matter from dwarf galaxies.

\subsubsection{Spatial Distribution of Satellites}

\begin{figure}
\epsfysize = 6cm \centreline{\epsfbox{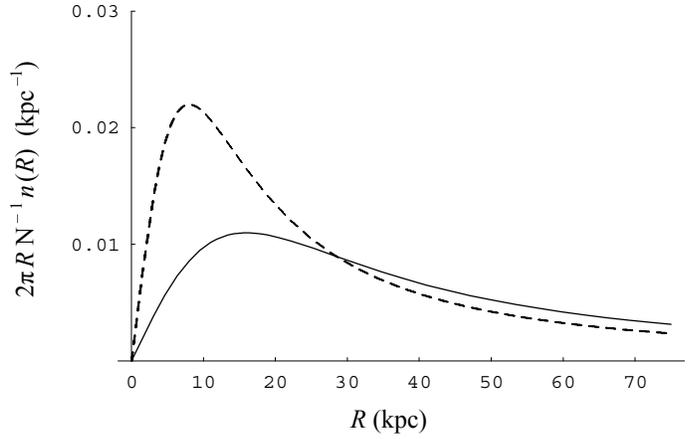}}
\caption{Probability density of satellites with respect to the
  lens-plane polar radius $R$.  The solid line is for $\Rfifty =
  100\kpc$, the dashed line for $\Rfifty = 50\kpc$.
\label{fig:SatellitePDF}}
\end{figure}

%
%
The distribution of satellite galaxies is taken as spherically
symmetric.  Explicitly, we assume that the number density in the lens
plane is
\begin{equation}\label{eq:2-Satellite-2D-density}
n(R) \propto {(R^{2} + \rd^{2})^{-m/2}} \:.
\end{equation}
where $m > 1$ determines the asymptotic fall-off of the distribution,
and $\rd$ is a central softening parameter.  We choose $m = 5/2$ so
that the three-dimensional number density, obtained by Abel
deprojection of \eqref{eq:2-Satellite-2D-density}, falls off like
${(\rm distance})^{-3.5}$, similar to the behaviour observed in the
Milky Way~\citep[see e.g.,][]{Wi99}.  We choose two different $\rd$,
so that half the satellites are within a (spherical polar) radius
$\Rfifty$ of $100\kpc$ or $50\kpc$ respectively of the centre of the
main galaxy. The distributions are plotted in
Fig.~\ref{fig:SatellitePDF}. Note that, at least as judged from the
case of the Milky Way, we expect half the satellites to lie within
$100$ kpc, and so $\Rfifty \sim 100 \kpc$ is perhaps the more
realistic.

\subsection{Numerical Details}

For a given source position, the lens equation can be solved
numerically by first triangulating the image plane, mapping the grid
to the source plane to triangulate the image positions, and then using
a multi-dimensional Newton-Raphson procedure.

The satellite galaxies have small effects on the positions of the
images and on critical curves and caustics. Their effect is mainly on
the ratios between image fluxes. When a satellite galaxy is
sufficiently near an image, it changes the magnification of that
image. Only if the separation between the satellite and the image is
still smaller does image splitting occur.  In this paper, we focus on
the effect of the satellites on the flux ratios of lenses rather than
the image multiplicity. When images split, they are near new critical
curves created by the satellite galaxy and are highly magnified, which
certainly results in a flux ratio change.

We generate positions of sources randomly to find 1000 five-image and
1000 three-image systems. For each system, the image positions and
magnifications are found numerically, both for the main galaxy alone
and with satellite galaxies at positions randomly generated according
to the distribution \eqref{eq:2-Satellite-2D-density} (or its 3D deprojection, for the purposes of finding tidal radii). We count
the number of systems where the satellite galaxies change the ratio of
any two image fluxes by $5\%$ or more (discounting the usually
unobservable central images).

\section{The Total Mass in Satellites}\label{sec:Total-mass-in-satellites}

\subsection{Plummer Satellites}\label{secsub:3-Plummer Satellites}

\begin{figure}
\epsfxsize = 8.38cm \centreline{\epsfbox{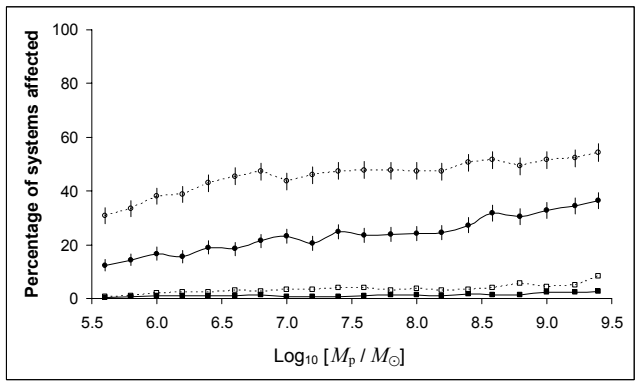}}
\caption{The proportions $F_{4}$ of quad (circles) and $F_{2}$ of doublet (squares) lens
  systems that have a flux ratio changed by $\geq 5\%$ by Plummer satellite
  galaxies, as a function of the mass per satellite galaxy. The total
  mass in satellites, $\Mdw$, is fixed at $5\E{9}\Msun$ and the scalelength
  of the distribution at $\Rfifty = 100\kpc$ (solid points and line) and $\Rfifty = 50\kpc$ (open points and dotted line). The lens is a
  pseudo-isothermal elliptic potential with $\epsilon = 0.1$.  Note
  especially that $F_{4}$ is only weakly dependent on how the mass is shared between satellites. The error bars
  are $2\sigma$ of a binomial distribution. \label{fig:CTM-Plummer} }
\end{figure}

For the moment, let us fix the total mass in satellite galaxies $\Mdw$
as $5\times10^{9}\Msun$ and share it equally among a varying number
$N$ of satellites.  The lens galaxy is an $\epsilon = 0.1$
pseudo-isothermal elliptical potential, and the satellites are
spatially distributed with $\Rfifty = 100\kpc$.  The results are shown
in Fig.~\ref{fig:CTM-Plummer}, which illustrates that, over a large
range of $\Mp$ or $N$, the proportion $F_{4}$ of quads with flux
ratios affected by the satellites is only weakly dependent on how mass
is apportioned between the satellites. That is, the proportion $F_{4}$
is mainly sensitive to the total mass $\Mdw = N\Mp$ rather than $N$ or
$\Mp$. The solid line in Fig.~\ref{fig:CTM-Plummer} suggests that
satellites of mass $10^{9}\Msun$ are only $50\%$ more efficient at
altering flux ratios than those of mass $10^{7}\Msun$.  The Plummer
scale radius $\rp$, if set differently, can affect the results. For
example, fixing $\rp$ to $140\pc$ for all $\Mp$ causes $F_{i}$ fall
off rapidly below $\Mp = 10^{7}\Msun$, as the central density of the
satellites decreases.

Fig.~\ref{fig:CTM-Plummer} can be understood qualitatively as
follows. The magnification of an image is changed by $\geq 5\% $ --
and the flux ratios of the lens system affected -- if the image is
within a `radius of influence' $\reff$ of a satellite. Here, $\reff$
depends on the location and magnification of the image, as well as the
mass of the satellite. ($\reff$ increases with the mass of the
satellite and the magnification of the image. Note that there can be
an ambiguity in $\reff$.  As a satellite moves closer to the image,
the image may be first slightly brightened, before dimming rapidly as
the satellite approaches, or first slightly dimmed before brightening
rapidly. So, for a large enough satellite, there may be two ranges in
which the flux changes by $\geq 5 \%$. For the purpose of our rough
argument here, however, we ignore this complication.) For a given
configuration, the probability that the flux ratio is changed
appreciably by satellites depends not only on the number of satellites
and their probability distribution in space, but on the individual
values of $\reff$. Even so, ensemble averaging over many image
configurations of the same type (e.g. quads) yields an $\reff$ that
depends only on the mass of the satellite.

The fractions of quads and doublets that are affected by satellites
depend on the fractional area $A_{\mathrm{sat}}$ of the lens plane
that lies within the circles of influence. That is,
\begin{eqnarray}\label{eq:3-F-function-AreaSat}
1 - F_{4} \approx (1 - A_{\mathrm{sat}})^{4} &\implies & F_{4} \approx 4 A_{\mathrm{sat}} - 6 A_{\mathrm{sat}}^{2}\;, \nonumber \\
1 - F_{2} \approx (1 - A_{\mathrm{sat}})^{2} &\implies & F_{2} \approx 2 A_{\mathrm{sat}} - A_{\mathrm{sat}}^{2}\;,
\end{eqnarray}
where $\reff, A_{\mathrm{sat}}$ are different for quads and doublets.
If $A_{\mathrm{sat}}$ is small, the circles of influence do not overlap, so
\begin{equation}\label{eq:3-Asat-Sum-piReffsqrd}
A_{\mathrm{sat}} \propto \sum_{i=1}^{N} \pi\reff^{(i) {2}} \:.
\end{equation}

The gently increasing $F_{i}$ in Fig.~\ref{fig:CTM-Plummer} over a
range of $\Mp$ reflects a dependence $ \reff^{2} \propto \Mp^{p}$ for
$p \gtrsim 1$. For $\Rfifty = 50\kpc$, the proportional increase in
$F_{4}$ between, say, $10^{7}\Msun$ and $10^{9}\Msun$ (see dotted line
in Fig.~\ref{fig:CTM-Plummer}) is less than for $\Rfifty = 100\kpc$
because of greater overlapping ($A_{\mathrm{sat}}$ increases more
slowly than $\sum_{i=1}^{N} \pi\reff^{(i) {2}}$). Doublets are much
less strongly affected than quads not only because there are only two
rather than four images whose fluxes could be affected (hence the $2
A_{\mathrm{sat}}$ rather than $4 A_{\mathrm{sat}}$ term in
\eqref{eq:3-F-function-AreaSat}), but, more importantly, because their
images are typically of much lower magnification.

\begin{figure}
\epsfxsize = 8.38cm \centreline{\epsfbox{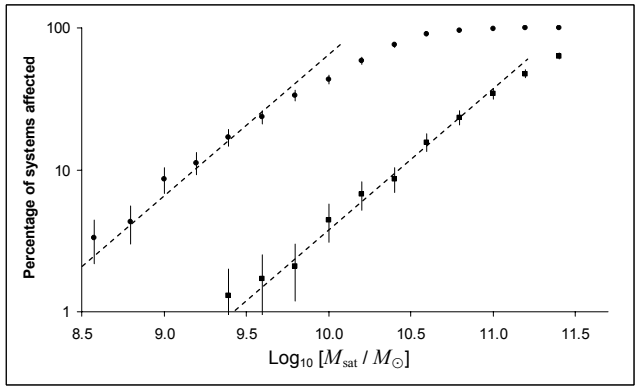}}
\caption{The proportions $F_{i}$ of quads and doublets (upper and
  lower curves) as functions of the total mass $\Mdw$ in Plummer satellites drawn from the $M^{-2}$ distribution, on a log-log
  scale. Here, $\Rfifty = 100\kpc;$ similar behaviour holds for other $\Rfifty$ (not illustrated). The dotted lines are of unit gradient: when the percentage of affected systems is low, $F_{i}$ scales as $\lesssim \Mdw$.\label{fig:LogPlot}}
\end{figure}

Let us now allow the total mass $\Mdw$ in satellite galaxies to
vary. Since $F_{i}$ are not quite independent of how $\Mdw$ is
apportioned between satellites, we need to allow for different
individual satellite masses. We draw them from an $M^{-2}$
distribution with cutoffs at $5.0\E{6}\Msun$ and $5.0\E{9}\Msun$, and
vary $N$ to vary $\Mdw$. (Cutoff masses of below $5.0\E{6}$ were too
computationally expensive because many more satellites would have been
needed for the same total masses.) The masses of the satellites (along
with their positions) are regenerated for each new source
position. The results are shown in Fig.~\ref{fig:LogPlot}, where
$F_{i}$ are plotted as a function $\Mp$ on a log-log scale.  The
scalings \eqref{eq:3-F-function-AreaSat} and
\eqref{eq:3-Asat-Sum-piReffsqrd} can be seen: $F_{i}$ initially
increase $\lesssim$linearly with $\Mdw$ (as $A_{\mathrm{sat}} \propto
N \propto \Mdw$ for a given satellite mass function), until the
overlaps between circles of influence can no longer be neglected,
after which $F_{i}$ approach unity asymptotically. The proportions
$F_{i}$ go from $20\%$ to $80\%$ over about one order of magnitude of
$\Mdw$ (this can be seen even more clearly from
Fig.~\ref{fig:Plummer-SixSets} in the next section). At all $\Mdw$,
flux ratios of quads are much more likely to be affected than those of
doublets.

\subsection{Hernquist Satellites}

\begin{figure}
\epsfxsize = 8.38cm \centreline{\epsfbox{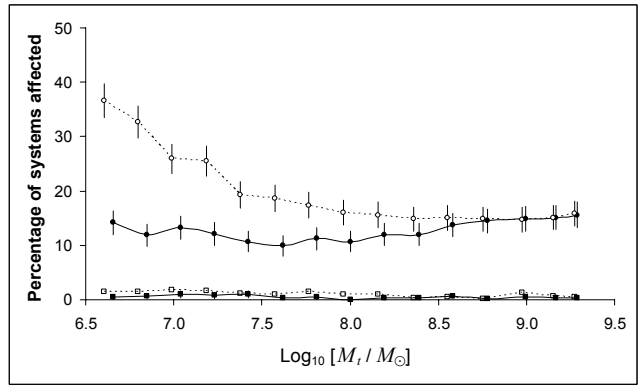}}
\caption{As Fig.~\ref{fig:CTM-Plummer}, but with an initial total mass
  in Hernquist profile satellites fixed at $1.0\E{10}\Msun$. $F_{i}$ are
  plotted against the average tidally bound mass per truncated
  Hernquist satellite.  \label{fig:CTM-SHQD}}
\end{figure}

Since the bound mass $\Mt$ of a Hernquist profile satellite depends,
through \eqref{eq:2-SHQ-Defn-of-rt} and \eqref{eq:2-SHQ-rs-Mt-scaling}
on its initial mass $M$ and its distance from the main galaxy
(generated randomly in this simulation), we cannot repeat
\S\ref{secsub:3-Plummer Satellites} exactly. We can only fix the total
initial mass, apportioning it equally between the varying number of
satellites $N$. More-massive satellites, which are therefore less
concentrated, are more prone to tidal disruption: for $\Rfifty =
100\kpc$, a total initial mass of $1.0\E{10}\Msun$ results in a total
tidally bound mass of $\sim7\E{9}\Msun$ if shared equally between 1000
dwarfs of initial mass $10^{7}\Msun$, and $\sim6\E{9}\Msun$ if shared
between 10 of initial mass $10^{9}\Msun$. For $\Rfifty = 50\kpc$ these
fall to $\sim 6.3\E{9}\Msun$ and $\sim 5.7\E{9}\Msun$
respectively. Even more significantly, the most massive satellites
cannot stay bound close-in to the main galaxy (whereas smaller
satellites \textit{can}), decreasing their chances of lying near the
centre of the lens in projection, where the images typically are. This
is dramatically illustrated in Fig.~\ref{fig:CTM-SHQD}, where for
$\Rfifty = 100\kpc$ the proportion of systems with altered flux ratios
only varies weakly with $N$, whereas for $\Rfifty = 50\kpc$ satellites
with $\Mt \lesssim 5\E{7}\Msun $ are much more efficient at changing
fluxes than massive ones. However, this result depends on the
assumption that a satellite which is highly tidally disrupted ($\rt <
\rs$) can be ignored, on the grounds that its mass surface density is
so diffuse that it has little effect on the lensing fluxes.

\begin{figure}
\epsfxsize = 8.38cm \centreline{\epsfbox{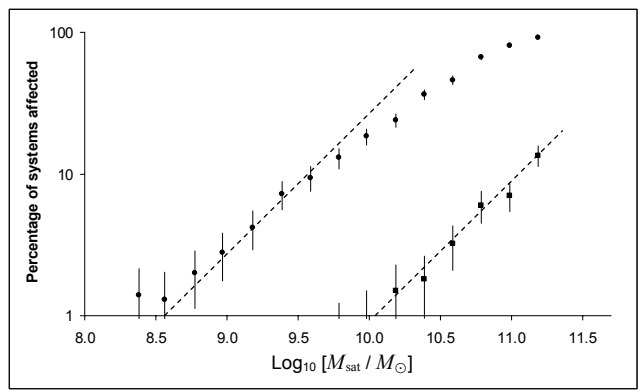}}
\caption{The proportions $F_{i}$ of quads and doublets (upper and
  lower curves) as functions of tidal mass $\Mdw$ in Hernquist dwarfs, $\Rfifty = 100\kpc$. The dotted lines are of unit gradient.\label{fig:LogPlot-SHQD}}
\end{figure}

Drawing initial masses from the same $M^{-2}$ distribution as before,
and varying $N$ to vary the total bound mass $\Mdw$, we obtain 
Fig.~\ref{fig:LogPlot-SHQD}, which is the analogue of
Fig.~\ref{fig:LogPlot}. The Hernquist profile satellites, which are
more extended and diffuse than the Plummer models, are less efficient
at altering fluxes.

\section{Astrophysical Applications}\label{sec:Astrophysical Applications}

\subsection{Elliptical Galaxy Lenses}

\begin{figure}
\epsfxsize = 8.38cm \centreline{\epsfbox{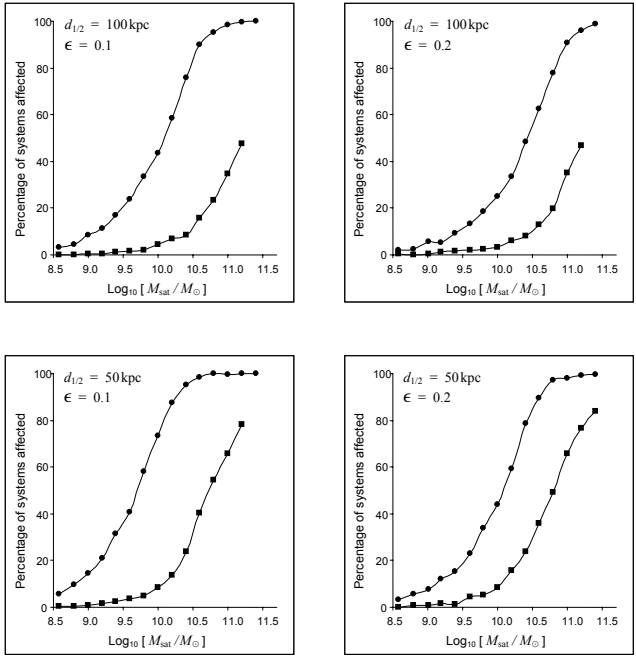}}
\caption{Plots of $F_{i}$ (circles for quads, squares for doublets) as
  functions of total mass $\Mdw$ in Plummer dwarfs, for various
  parameters. The lens is a pseudo-isothermal elliptic potential with
  ellipticity $\epsilon$, while the satellite distribution has a
  characteristic scalelength of $\Rfifty$ as indicated in the top
  left of each panel. \label{fig:Plummer-SixSets}}
\end{figure}


Let us consider several different sets of parameters of elliptical
potentials~(\ref{eq:2-DefPIEP}) and ask what mass in satellites is
required for there to be a significant probability of flux-ratio
changes. We first use Plummer model satellites, drawn from the
$M^{-2}$ mass function, and again vary the number of satellites to
vary the total mass. The results are plotted in
Figure~\ref{fig:Plummer-SixSets}.

For the moderately elliptic case $\epsilon = 0.1$, masses of $\sim
1.25\E{10} \Msun$ are needed for 50\% of quadruplets to show anomalous
flux ratios if $ \Rfifty = 100\kpc, $ decreasing to $\sim 5\E{9}$ for
$\Rfifty = 50\kpc$. The
lengthscale of the satellite distribution $\Rfifty$ has a large effect
on \textit{both} $F_{4}$ and $F_{2}$. As expected, the higher the
probability density of satellites in the inner parts of the lens plane
(where the images lie), the greater the proportion of anomalous flux
ratio systems. Comparing left and right panels, we see that the effect
of satellites on flux ratios of quadruplets decreases as $\epsilon$
increases -- and the quadruplet cross-section increases, while the
mean magnification of quads falls. The proportion of doublets $F_{2}$,
however, is \textit{not} noticeably affected by the ellipticity
$\epsilon .$

\begin{figure}
\epsfxsize = 8.38cm \centreline{\epsfbox{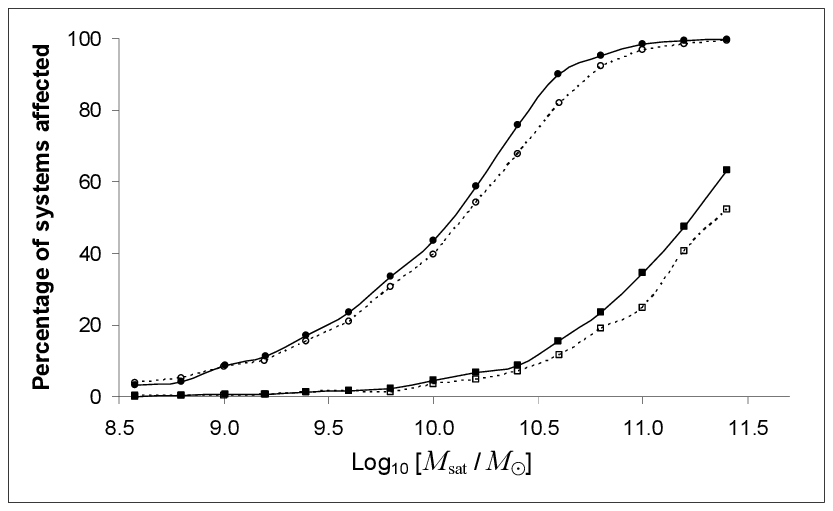}}
\caption{Proportion of systems with flux ratios affected by Plummer
  model satellite galaxies, as a function of total mass in satellites,
  for two sets of redshifts. Filled points and solid line: $\zl =
  0.46$, $\zs = 2.15$ (corresponding to $\Sigmacrit = 2.04\E{9}
  \Msun\kpc^{-2}$). Open points and dotted line: $\zl = 0.21$, $\zs =
  1.22$ (corresponding to $\Sigmacrit = 3.05\E{9}
  \Msun\kpc^{-2}$). The first set is the median lens- and
  source-redshifts of known 4-image systems; the second $\zl$ and
  $\zs$ are one standard deviation below the
  median. \label{fig:Indep-of-Redshift}}
\end{figure}

\begin{figure}
\epsfxsize = 8.38cm \centreline{\epsfbox{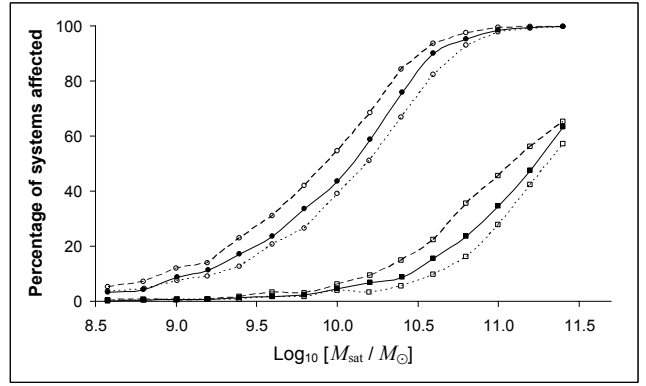}}
\caption{Percentages of systems with flux ratios affected by Plummer
  model satellite galaxies, as a function of the total mass in
  satellites, for redshifts $\zl = 0.46$ and $\zs = 2.15,$ for three
  different velocity dispersions: $\sigma = 200\kmpersec$ (dashed
  line), $250\kmpersec$ (solid line), $300\kmpersec$ (dotted
  line). The remaining parameters are $\epsilon = 0.1$ and $\Rfifty =
  100\kpc$. \label{fig:Dep-on-VelDisp}}
\end{figure}

Changing the redshifts of lens and source affects these results only
through $\Sigmacrit$, on which $\Er$ in eq~(\ref{eq:wyn}) and
$\kappa_0^{(k)}$ in eq~(\ref{eq:2-DefPlummer-kappa}) depend. $F_{i}$
for two sets of redshifts are plotted in
Fig.~\ref{fig:Indep-of-Redshift}, for $\Rfifty = 50\kpc$ and $\epsilon
= 0.1$. We see that the results are not very sensitive to different
redshifts or different $\Sigmacrit$.
The results are also not very sensitive to the core radius of the
pseudo-isothermal elliptic potential, although they \textit{are}
affected by its velocity dispersion $\sigma$, as shown in
Fig.~\ref{fig:Dep-on-VelDisp}. We recall that the velocity dispersion
controls the Einstein radius $\Er = \sigma^{2}(\xi_{0} G
\Sigmacrit)^{-1} $, with more massive lens galaxies diluting the
effect of the satellites on flux ratios. A pseudo-isothermal elliptic
potential with $\sigma = 300\kmpersec$ requires $\sim 2$ times as much
mass in satellite galaxies as a $\sigma = 200\kmpersec$ model for
$F_{4} = 0.5$, or 50\% of the quads to have anomalous flux
ratios. However, the Einstein radius of the main galaxy goes as
$\sigma^{2}$, and the projected mass within the Einstein ring as
$\sigma^{4}$, so a given \textit{relative} mass in satellite
companions is more likely to affect flux ratios in more massive
elliptical lenses.

\begin{figure}
\epsfxsize = 8.38cm \centreline{\epsfbox{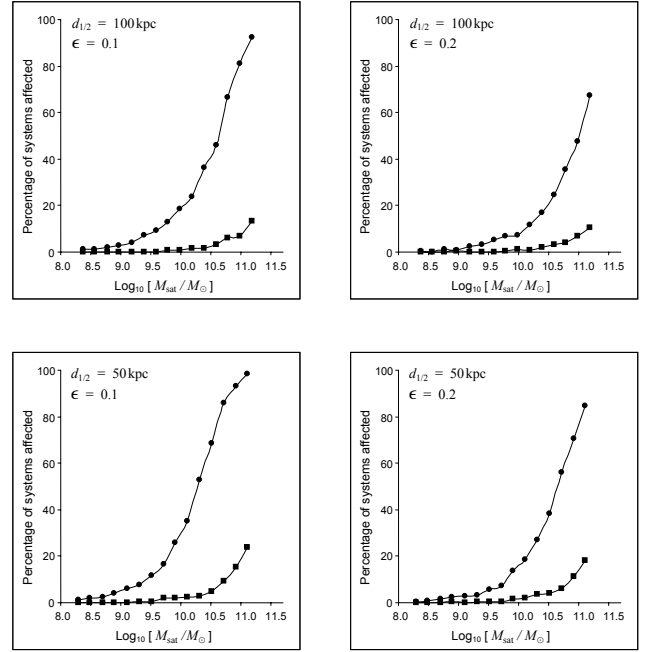}}
\caption{Effect of Hernquist satellites on flux ratios (analogue to 
Fig.~\ref{fig:Plummer-SixSets}). \label{fig:SHQD-FourSets}}
\end{figure}

When the less concentrated, tidally-stripped Hernquist profile
satellites are used instead of Plummer models, the effect on flux
ratios is significantly weaker: Fig.~\ref{fig:SHQD-FourSets} is the
analogue of Fig.~\ref{fig:Plummer-SixSets}. The initial masses of the
Hernquist dwarfs are drawn from the same $M^{-2}$ distribution, and
$F_{i}$ are plotted against the total tidal mass in satellites. For
the $\epsilon = 0.1$ elliptical potential, if $\Rfifty = 100\kpc$,
some $4.5\E{10}\Msun$ in satellites is needed for $50\%$ of quadruplet
flux ratios to be altered (over three times the mass in Plummer
satellites needed), dropping only to $2\E{10}$ for $\Rfifty = 50\kpc$
(about eight times the mass in Plummer satellites needed) because of
the increased tidal stripping of closer-in satellites.

\subsection{Spiral Galaxy Lenses}\label{sec:Spirals}

\begin{figure}
\epsfxsize = 8.38cm \centreline{\epsfbox{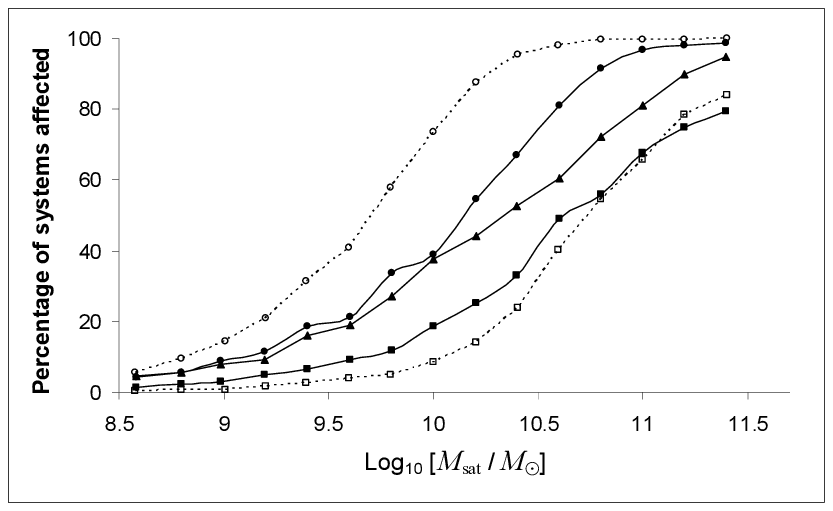}}
\epsfxsize = 8.38cm \centreline{\epsfbox{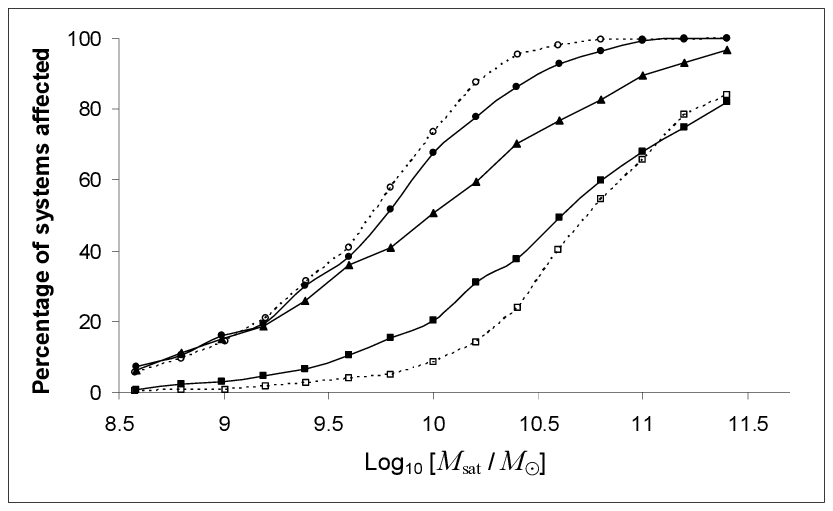}}
\caption{The upper panel shows $F_{i}$ as functions of the total mass
  in Plummer satellites, for an edge-on Milky Way with maximal disc (solid
  lines and filled symbols) and a pseudo-isothermal elliptic potential
  for comparison (dotted lines and open symbols) with $\sigma =
  250\kmpersec$. The lower panel shows the same quantities but for a
  sub-maximal disc.  Both galaxies are at $\zl = 0.46,$ lensing
  sources at $\zs = 2.15.$ $\Rfifty = 50\kpc.$ Quads are shown as
  circles, doublets as squares, and disc triplets as
  triangles. \label{fig:MW-versus-PIEP}}
\end{figure}

Now let us change the lens galaxy to an edge-on spiral, using the
model based on
eqns~(\ref{eq:MW-HernquistDensity})-(\ref{eq:MW-MN-DefPot}).  Plots of
$F_{i}$ as functions of $\Mdw$, for Plummer model satellites, are
plotted in Fig.~\ref{fig:MW-versus-PIEP} as solid lines alongside
those for a pseudo-isothermal elliptic potential as dotted lines. The
upper panel shows the model in which the disc is maximal and provides
most of the rotational support in the inner parts, the lower panel
shows a model in which the disc is sub-maximal (see \citealt{Sh07} for a
detailed discussion of the lensing properties of these models). There
are three solid lines in the panels, as the results are divided
according to the five image (in effect, quadruplet), core triplet (in
effect, doublet) and disc triplet morphologies.

Compared to a typical elliptical, the flux ratios of maximal disc lens
quads are significantly less susceptible to changes through
substructure -- almost three times the mass in satellite galaxies is
needed to affect half the flux ratios. Spiral doublets, however, are
slightly more susceptible. This is as expected: the greater the
asymmetry of the matter distribution, the larger the 4-image cross-section,
the lower the typical magnification of quads and the smaller the
effect of satellite galaxies on the 4-image flux ratios. (So disc
triplets, with typical magnifications in between quads and doublets,
are also in between in susceptibility to flux changes.)  Indeed, we do
not expect many anomalous flux ratios in disc galaxy lenses to be
caused by the satellite galaxies in their haloes unless these
satellites total at least $\gtrsim 10^{10}\Msun$ in mass, that is,
$\approx 10\%$ of the luminous galaxy mass! (And \textit{that} assumes
the satellites are distributed with $\Rfifty = 50\kpc$.) On moving to
the sub-maximal case, the 4-image and disc triplet cross sections are
substantially reduced, and so the numbers return towards their values
in the elliptical galaxy case.

Note that we have compared the effects of a typical spiral with
asymptotic circular speed of $v_0 \approx 220 \kmpersec$ to a typical
elliptical with line-of-sight velocity dispersion of $\sigma = 250
\kmpersec$. This makes sense, as early-type galaxies are more massive
than late-type. If instead we were to carry out the comparison using
$\sigma = v_0/\sqrt{2}$, so that the kinematic properties of the
models were comparable, then the elliptical would have a lower
velocity dispersion and so the discrepancy between the effects of
spirals and ellipticals would be increased (see
Figure~\ref{fig:Dep-on-VelDisp}).

\subsection{Visible Substructure}

\begin{table*}
\begin{center}
\caption{The proportion $F_{4}$ of quads with anomalous flux ratios
  for five runs of the two-population model. Here, the satellite
  populations have $\Rfifty = 100\kpc$ and the elliptical lensing
  galaxy has the standard parameters. Population 1 satellites have
  mass $1.0\E{7} \Msun$, while population 2 satellites have mass
  $1.0\E{9} \Msun$. (For Hernquist satellites, these are
  \textit{initial} masses.) The Plummer model Population 2 contributes
  to $F_{4}$ by $\sim50\%$ more than population 1 satellites, whilst
  the two Hernquist-model populations contribute equally.
\label{tab:edone}}
\begin{tabular}{cccccccc}
\hline
Satellite   & $N_{1}M_{1}$  & $N_{2}M_{2}$ & $F_{4}$ & $F_{4,1}$ & $F_{4,2}$   & $N_1M_1:N_2M_2$ & $F_{4,1}:F_{4,2}$   \\
Profile    & $(\Msun)$     & $(\Msun)$   & (Both pops)  & (Pop 1 only)  & (Pop 2 only) &  & \\
\hline
Plummer    & $1.0\E{10}$ & $1.0\E{9}$ & 43.2\% & 37.3\% & 12.2\% & $10:1$& $3.1:1$\\
Plummer    & $5.0\E{9}$  & $5.0\E{9}$ & 45.6\% & 21.5\% & 34.5\% & $1:1$ & $1:1.6$\\
Plummer    & $6.0\E{9}$  & $4.0\E{9}$ & 44.3\% & 23.4\% & 28.1\% & $3:2$ & $1:1.2$\\
Hernquist  & $2.0\E{10}$ & $2.0\E{9}$ &  28.4\% &  25.1\% &  5.1\% & $10:1$ & $4.9:1$ \\
Hernquist  & $1.2\E{10}$ & $8.0\E{9}$ &  22.8\% &  16.3\% &  10.6\% & $3:2$ & $3.1:2$ \\
Hernquist  & $1.0\E{10}$  & $1.0\E{10}$ & 20.9\% & 13.2\% & 12.1\% & $1:1$ & $1.1:1$ \\
\hline
\end{tabular}
\end{center}
\end{table*}

Some flux anomalies appear to be attributable to single dwarf
galaxies. The most obvious examples are the Sixth Object in
MG0414+0534~\citep{Sc93} and G2 in CLASS~B2045+265 ~\citep{Mc07}. In
both these cases, a single piece of substructure gives a substantial
improvement in the fit of a smooth model of the lens galaxy.  An
interesting question is -- if flux ratio anomalies are due to
substructure -- how often might we expect to see a visible
counterpart? At the typical redshifts of the lenses, only the very
largest dwarf galaxies can be detected with ground-based telescopes,
of course.

This question can be answered by simulations in which the satellites
are divided into two populations -- $N_{1}$ dark satellites of low
mass $M_{1}$ and $N_{2}$ bright satellites of high mass $M_{2} \gg
M_{1}$ -- and flux-ratio changes are sought for the two populations
separately and together. Some sample results of the simulations are
shown in Table~\ref{tab:edone} for the usual $\epsilon = 0.1$
pseudo-isothermal elliptic potential with $\Rfifty = 100\kpc$. Note
that, for the Plummer satellites, when the ratio of the total mass in
population 1 to total mass in population 2 is 1:1, then the number of
systems with anomalous flux ratios caused by population 1 as compared
to population 2 is roughly in the ratio of 2:3, and a 3:2 mass ratio
gives a roughly 1:1 effect. For the Hernquist satellites, a 1:1 mass
ratio gives 1:1 effect. This confirms the observations of
\S\ref{sec:Total-mass-in-satellites} that a $10^{9}\Msun$
Plummer-model dwarf spheroidal is only $\sim50\%$ more efficient at
perturbing fluxes than a $10^{7}\Msun$ one, and more-massive Hernquist
dwarfs are \textit{no} more efficient than less massive ones. So from
Table~\ref{tab:edone}, we see that the most massive satellites do not
contribute very disproportionately to anomalous flux ratios.

The neat correspondence between ratios is not seen for 10:1 mass
ratios because $F_{4}$ increases sub-linearly with mass: e.g. a single
$10^9\Msun$ in Plummer satellite changes fluxes in $12\%$ of quads,
but four of them change fluxes in $28\%$, not $48\%$, of
quads. However, it is confirmed that $\sim 85\%$ of systems with
affected fluxes still have affected fluxes if the $10\%$ of mass in
massive satellites is removed.

The fraction of observed anomalous flux ratio systems with visible
(and therefore high mass) culprits is actually quite high. Of the six
four-image lens systems proposed by \citet{Da02}, three (B2045+265,
MG0414+0534, B1608+656) have identified, visible substructure that may
cause the flux perturbation. If this datum is taken at face value, it
suggests that about half the mass in substructure is in dwarf galaxies
large enough to be optically identified. Simulations, however, tend to
find that the satellite masses behave more like an $M^{-2}$
distribution. This implies that there is equal mass in equal decades,
and therefore that each decade is responsible for the causing roughly
the same number of flux ratio anomalies (the $10^{9}\Msun$ decade
causing only $50\%$ more than the $10^{7}\Msun$ decade).

One resolution of this difficulty is to postulate that the visible
dwarf has been mistakenly designated as the culprit and that the
anomalous flux ratio is really produced by another dark satellite. A
hint that this may sometimes be the case is given by the
unrealistically large flattening deduced for satellite G2 in CLASS
B2045+265 using model-fitting in \citet{Mc07}. In other words, it may
simply be a chance effect that many anomalous flux ratio systems
appear to have visible objects at or near the Einstein radius.  The
probability that a large dwarf lies close to the Einstein radius is
easily computed from eq~(\ref{eq:2-Satellite-2D-density}). When
$\Rfifty = 100\kpc$, then there is a $10\%$ probability of a large
dwarf lying within two Einstein radii and $3\%$ within one. This still
does not seem large enough to explain the effect, but caution is
needed as there may be other supplies of substructure along the line
of sight for some of the lenses in groups and clusters, like
B1359+154. It is worth noting that simulations (see e.g. \citealp{Ze05}) can sometimes yield
highly anisotropic distribution of substructure in simulated halos: the projected subhalo mass within $10$ kpc can vary by factor of 10 depending on viewing angle. In this case, sightlines which project
massive satellites onto small radius are also much more likely to
project other satellites onto small radius.  This may mean that our
computed probabilities of 3-10 \% may be on the low side.

Another possible resolution of the difficulty is that luminous
satellites, the baryons in which have cooled and condensed, may be much
more centrally concentrated than dark satellites.  In other words, it
is possible that luminous satellites would have a much larger effect
on flux ratios than their dark brethren because they are structurally
different and more compact. Of course, the effect of baryons on dark
haloes is subject to considerable uncertainties.  This effect must be
small for dwarfs like Draco, with a mass-to-light ratio of $>
350$~\citep{Kl01}, although it may be more significant for satellites
like the Large Magellanic Cloud with a mass-to-light ratio of $\sim
5$~\citep{Al04}.  It is also known that semianalytic calculations of
galaxy formation lead to too many compact, luminous satellites, as
compared to what is seen around the Milky Way (see e.g., Koposov et
al. 2007). The bright satellites predicted by semi-analytic theories
are much too concentrated, suggesting that this effect is overplayed
in the modelling.

Nonetheless, the effect certainly exists at some level and is worth
investigating. Baryon condensation may increase the central density
by a factor of between 4 and 160~\citep{Gn02}, although the larger
numbers are probably more appropriate for galaxies like the Milky Way
rather than satellite galaxies. We change the lengthscale $\rs$ of the
Hernquist model to mimic the result of baryon condensation and
quantify the extra effect, compared to dark Hernquist satellites, that
highly-concentrated luminous satellites could have on flux
ratios. (The Plummer satellites are already so compact that changing
the Plummer scale radius makes no appreciable difference even when the
central density is raised by a factor 100.) The Hernquist density law
\eqref{eq:2-SHQ-rho} means that the central density goes as
$\rs^{-2}$, so we modify the scalelength-to-mass relation
\eqref{eq:2-SHQ-rs-Mt-scaling} to
\begin{equation}\label{eq:4-SHQ-rs-Mt-qfac-scaling}
\rs = q \; 10^{-4} \sqrt{\Mt / \Msun} \:,
\end{equation}
where $q = 1,\; 1/\sqrt{10},\;\text{or} 0.1$, corresponding to central
densities of 1, 10 and 100 times that of the dark (uncontracted)
Hernquist profile. The number of satellites of initial mass $10^9
\Msun$ needed to affect flux ratios in $50\%$ of four-image
systems (all other parameters being the same as in
Table~\ref{tab:edone}) is shown in Table~\ref{tab:50pc-qfactor} .
\begin{table}
\begin{center}
\caption{The (average) tidal mass $\Msat$ in Hernquist satellites of
untruncated mass $10^{9}\Msun$ needed to affect fluxes in $50\%$ of
four-image systems, for various $q$. The elliptical galaxy has the
standard parameters and the scalelength of the spatial distribution of
satellites is $\Rfifty = 100\kpc$.
  \label{tab:50pc-qfactor}}
\begin{tabular}{cccc}
\hline
Satellite   & $N$  & $\Msat$ & \\
Profile    &       & $(\Msun)$ \\
\hline
Hernquist (dark)  & $100$ & $5.6\E{10}$ \\
Hernquist ($10\times$ concentration) & $20$ & $1.5\E{10}$ \\
Hernquist ($100\times$ concentration) & $12$  & $1.1\E{10}$ \\
\hline
\end{tabular}
\end{center}
\end{table}
Almost four times the tidal mass in massive dark satellites is
required, compared to massive luminous satellites with $10$ times the
central density, to affect the same proportion of flux ratios. Raising
the central density a further factor of 10 has a smaller effect. The
increasing effect on fluxes seen with decreasing $\rs$ is amplified by
the extra resistance to tidal disruption of the more-concentrated
satellites.

Table~\ref{tab:2popqfac} shows the results of two-population models,
in which population 1 are dark satellites of initial mass $10^{7}\Msun$ and population 2 are bright
satellites of initial mass $10^{9}\Msun$, with 10 or 100 times the central density of their dark
brethren. The fraction of anomalous flux ratio systems caused by
bright substructure is now impressively high -- for example, the
second line of the table tells us that 42 \% of systems have anomalous
flux ratios, of which 28 \% remain anomalous when the dark population
is removed. In other words, over half of the anomalous flux ratio
systems are caused, at least in part, by bright satellites. This is
close to the statistics on observed systems -- although caution is
needed as large compression factors like 10 or 100 may well cause the
importance of this effect to be overestimated for satellite galaxies.

\begin{table*}
\begin{center}
\caption{The proportion $F_{4}$ of quads with anomalous flux ratios
  for three runs of the two population model, with the same parameters
  as in Table~\ref{tab:edone} but where population 2 satellites
  have increased central density. As the compression increases ($q$
  decreases), the tidally bound mass of population 2 satellites
  increases, but their effect on fluxes increases disproportionately,
  outstripping the effect of the diffuse low-mass population 1
  satellites. The difference caused by a factor 10 increase in central
  density is much larger than the difference caused by an extra step
  to a 100-fold increase.
\label{tab:2popqfac}}
\begin{tabular}{cccccccc}
\hline
Satellite   & $N_{1}$  & $N_{2}$ & $\Msat^{\mathrm{pop 1}}$ & $\Msat^{\mathrm{pop 2}}$ & $F_{4}$ & $F_{4,1}$ & $F_{4,2}$ \\
Profile    & $(\Msun)$     & $(\Msun)$  & $(\Msun)$ & $(\Msun)$  & (Both pops)  & (Pop 1 only)  & (Pop 2 only)  \\
\hline
Hernquist  & $1200$ & $8$ & $7.4\E{9}$ & $4.5\E{9}$ & 23\% &  16\% &  11\% \\
Hernquist (Pop 2 with $10\times$ concentration)  & $1200$ & $8$ &  $7.4\E{9}$ & $6.0\E{9}$ & 42\% &  16\% &  28\% \\
Hernquist (Pop 2 with $100\times$ concentration) & $1200$  & $8$ & $7.4\E{9}$ & $7.2\E{9}$ & 48\% & 16\% & 36\%  \\
\hline
\end{tabular}
\end{center}
\end{table*}

\section{Conclusions}

It remains unclear whether dark matter satellites and substructure are
responsible for anomalous flux ratios in strong lensing. \citet{Da02}
originally studied a sample of 7 radio-loud four-image lens systems
and claimed evidence of anomalies in 6 of them. This seemingly
suggests that anomalous flux ratios are very common. Here, we have
carried out a theoretical study of the frequency of flux ratio
anomalies as a function of lensing galaxy and dark matter substructure
parameters.

The likelihood that satellites affect flux ratios in strong lenses
depends on their mass profile. Here, we considered compact Plummer
spheres and diffuse, tidally stripped Hernquist profiles for our
satellites, with the satellite size scaling as a power of mass.  As
the Hernquist satellites are more extended than the Plummer models,
they therefore have a smaller effect on image fluxes -- typically
about 3 times the mass is needed to generate the same numbers of
anomalous flux ratios.

The probability that strong lensing flux ratios are affected by
satellites is crucially dependent on their spatial distribution. The
characteristic lengthscale $\Rfifty$ has a large effect on the
fraction of lenses with affected flux ratios.  Our spatial
distributions of satellite galaxies are inspired by the observational
data on the Milky Way, for which the satellite number density falls
off as $\rhat^{-3.5}$ in three-dimensions with a lengthscale of
$\Rfifty \sim 100 $ kpc. For such distributions, most satellites are
too far out to affect the fluxes of images. For example, even with
Plummer satellites, a mass of $\sim 3\E{9} \Msun$ is needed for 20 \%
of quadruplets to show anomalous flux ratios for a typical elliptical
galaxy, rising to $\sim 1.25 \E{10} \Msun$ for 50 \%.  Lenses that are
edge-on spiral galaxies with maximum discs (like the Milky Way) are
more resistant to flux changes by satellites, so the mass in
satellites and substructure has to be roughly a factor of 3 times as
great for the same proportion of quads to be affected. To obtain
anything like the apparent abundance of anomalous flux ratios, then
the scalelength of the substructure has to be different to what is
known for the Milky Way satellites.

Whether the flux ratios in a lens system are affected by
satellites is sensitive to the total mass in satellites, but more
weakly dependent of how this mass is apportioned between them. For
Plummer model satellites, the probability that a given satellite
changes a flux ratio increases with its mass only slightly faster than
linearly, at least when its mass is between $\sim5\times10^{6}\Msun$ and
$\sim10^{9} \Msun$. For example, satellites of mass
$\sim10^{9}\Msun$ are only responsible for the causing $\sim50\%$
more flux ratio anomalies than those of mass $\sim10^{7}\Msun$. For
Hernquist model satellites, more massive ones seem no more efficient
(per unit mass) at changing fluxes; indeed, more massive dwarfs, being
more prone to tidal disruption, might even be \textit{less} efficient
than lighter, more compact ones. One interesting consequence is that,
if matter in dark satellites is not predominantly in the most massive
ones, then the contribution of the most massive satellite galaxies to
flux ratio anomalies should not be predominant.

In the light of this, the fact that so many anomalous flux ratios
systems have optically identified substructure seems at outset
surprising. There seem to be two possible explanations. First, the
visible substructure may have been wrongly identified as the cause,
whereas dark substructure may be the true culprit. A large dwarf
galaxy may by chance be projected close to the Einstein radius,
whereas unrelated dark substructure may be the major cause of the
anomaly.  Second, visible satellites may be more concentrated than
their dark cousins, a physical effect that may naturally arise from
baryon condensation. Compression factors causing an enhancement of the
central density by a factor of 10 in bright satellites seem to be
ample to give a satisfactory explanation of the observed statistics.
Nonetheless, such high compression factors are probably implausible
except for the largest satellite galaxies. This seems to be in accord
with the results of \citet{Ma07}, who found that including baryons in
numerical simulations did not help in reconciling simulation results
with the statistics of anomalous flux ratios.

Finally, we remark that the likelihood that flux ratios are affected
depends on the ellipticity of the main lens galaxy, but this
dependence is much stronger in quads than doublets. This is a
consequence of high magnification images being more easily affected by
a dwarf galaxy than low magnification ones. Quads are more highly
magnified than doublets (the disc triplets of spirals are in between),
and changing the ellipticity of the main galaxy changes the typical
magnification of quads more than it changes that of
doublets. Generally, the greater the ellipticity the less the effect
of satellite galaxies. For example, a given mass $\Mdw$ of dwarf
satellites around a typical edge-on maximum-disc spiral galaxy is
significantly less likely to change image flux ratios than $\Mdw$
around a typical elliptical galaxy, even though the spiral is less
massive than the elliptical.

\section*{acknowledgements}
We thank the referee for some helpful comments. EMS thanks the
Commonwealth Scholarship Commission and the Cambridge Commonwealth
Trust for the award of a Studentship. NWE thanks Neal Jackson for some
insightful discussions. This work has been supported by the ANGLES
network.



\label{lastpage}

\end{document}